\documentclass[prl,twocolumn,showpacs,english,10pt,floatfix,groupedaddress,nofootinbib]{revtex4}
\usepackage{graphicx}
\usepackage{amssymb}
\usepackage{amsmath}

\usepackage[T1]{fontenc}
\usepackage[latin9]{inputenc}
\usepackage{color}
\usepackage{babel}

\usepackage{esint}

\makeatletter

\makeatletter
\usepackage{pst-node}\usepackage{lscape}\usepackage{amsfonts}\usepackage{amsbsy}\usepackage{graphics}\usepackage{lscape}\usepackage{mathrsfs}\usepackage{dsfont}

\newcommand{\qb}{\bar{q}}
\newcommand{\kv}{\vec{k}}

\renewcommand{\d}{\mathrm{d}}
\renewcommand{\i}{\mathrm{i}}

\newcommand{\bea}{\begin{eqnarray}}
\newcommand{\eea}{\end{eqnarray}}

\def\slash#1{\setbox0=\hbox{$#1$}  
   \dimen0=\wd0     
   \setbox1=\hbox{/} \dimen1=\wd1  
   \ifdim\dimen0>\dimen1   
      \rlap{\hbox to \dimen0{\hfil/\hfil}} 
      #1     
   \else     
      \rlap{\hbox to \dimen1{\hfil$#1$\hfil}} 
      /      
   \fi}      %

\usepackage{babel}

\makeatother

\begin{document}

\title{Probing Gluonic Spin-Orbit Correlations in Photon Pair Production}

\author{Jian-Wei Qiu$^{1,2}$, Marc Schlegel$^3$ and Werner Vogelsang$^3$}
\affiliation{$^1$Physics Department,
                  Brookhaven National Laboratory,
                  Upton, New York 11973, USA}
\affiliation{$^2$C.N. Yang Institute for Theoretical Physics,
                  Stony Brook University,
                  Stony Brook, New York 11794, USA}
\affiliation{$^3$Institute for Theoretical Physics, T\"{u}bingen University, 
                  Auf der Morgenstelle 14,
                  D-72076 T\"{u}bingen, Germany}

\date{\today}
\begin{abstract}
We consider photon pair production in hadronic collisions at large mass and 
small transverse momentum of the pair, assuming that
factorization in terms of transverse-momentum dependent 
parton distributions applies. The unpolarized cross section is found 
to have azimuthal angular dependencies that are generated by a gluonic 
version of the Boer-Mulders function. In addition, the single-transversely 
polarized cross section is sensitive to the gluon Sivers function. We 
present simple numerical estimates for the Boer-Mulders and Sivers 
effects in diphoton production at RHIC and find that the process 
would offer unique opportunities for exploring transverse-momentum 
dependent gluon distributions.
\end{abstract}

\maketitle

%
%
\noindent
{\it{ Introduction.}}\,---\,Hard hadronic processes with 
small transverse momentum $q_T$ of an observed
final-state system have attracted a lot of interest
because of their sensitivity to intrinsic parton transverse momenta. 
Such processes
may hence offer detailed insights into the partonic substructure of hadrons,
in terms of transverse-momentum dependent parton distributions (TMDs). Of
particular interest are correlations of the parton transverse 
momentum with the nucleon or quark spin, which are expressed by the
Sivers~\cite{sivers} and Boer-Mulders (BM)~\cite{bm} functions. From these, 
one ultimately hopes to learn about spin-orbit correlations and orbital 
angular momenta of partons confined in a nucleon. 
So far, the main focus of the field has been on quark TMDs. This is due to the 
fact that quark TMDs are primarily probed in semi-inclusive 
deep-inelastic 
scattering (SIDIS) and the Drell-Yan (DY) dilepton production process,
which have been accessible experimentally~\cite{Barone:2010zz}. On the 
theoretical side, the relative simplicity of these two reactions has allowed 
to derive factorization theorems involving TMDs~\cite{Collins:1981uk,tmdfact}.  

Gluon TMDs~\cite{rodrig} and processes sensitive to them have received 
closer attention only quite recently, at least for cases where
nucleon or gluon polarization matter. Several processes for accessing
the Sivers~\cite{boerwv,gluoncharm} gluon distribution, or the
gluonic version of the BM function (more appropriately 
described as the TMD distribution of linearly polarized gluons in an 
unpolarized nucleon)~\cite{Boer:2010zf,Boer:2009nc}, have been 
proposed 
for high-energy hadronic collisions, in particular, at the Relativistic Heavy 
Ion Collider (RHIC), or for $ep$ scattering at a future Electron 
Ion Collider (EIC). There is a generic dilemma concerning the 
processes considered so far: in cases where experiments can be carried
out at today's hadron colliders, factorization is known to be broken for
TMDs~\cite{rogers}, or to hold at best for weighted asymmetries that
only give information on 
integrated TMDs with certain transverse momentum weights.
On the
other hand, while transverse-momentum dependent factorization is expected to 
hold for reactions such as $ep\to c\bar{c}X$ or $ep\to {\mathrm{jet}}\,
{\mathrm{jet}}X$~\cite{Boer:2010zf}, realization of an EIC is still
a decade or so away. 

In this Letter we argue that the process $pp\to\gamma\gamma X$ can be used to 
study spin-dependent gluonic TMDs in a theoretically clean process already 
at RHIC. In proposing this process,
we are motivated by the following observations. First of all, since the 
final state is a color singlet, the diphoton process is expected 
to share many features with the DY process, as far as factorization is concerned. 
Indeed, like the DY process, its lowest-order contribution comes from $q\bar{q}$ 
annihilation, $q\bar{q}\to \gamma\gamma$, which can be shown to give rise 
to the same Wilson lines as the DY subprocess $q\bar{q}\to \gamma^\ast$, 
and hence involves the same quark and antiquark TMDs. Second, it has been
known for a long time that in the spin-averaged case~\cite{Berger} 
at colliders photon pair production is in fact dominated by the
process $gg\to \gamma \gamma$, that is, gluon-gluon 
fusion to a photon pair via a quark box. Even though this process is
formally down by two powers of the strong coupling constant $\alpha_s$
with respect to $q\bar{q}\to \gamma\gamma$, the suppression is
compensated by the structure of the associated hard-scattering
function, and by the size of the gluon distribution function.  
Hence, an experimental study of gluon TMDs should in principle be 
possible in this process. Finally, in order to study TMDs, precise 
measurement of the (small) transverse momentum of a final state is 
crucial. It seems to us that this should be easier to achieve for a
photon pair than, for example, for the jet pair in the reaction
 $ep\to {\mathrm{jet}} {\mathrm{jet}}X$. 

Being a background to a possible Higgs boson decay into two photons, 
QCD diphoton production has received a lot of attention in theoretical 
studies, in particular, for the diphoton pair transverse-momentum distributions
based on perturbative all-order resummation of Sudakov logarithms ~\cite{Nadolsky,Catani}.
In fact, these studies pointed out that the resummation 
formalism naturally suggests the presence of gluon TMDs, among 
them a perturbative spin-flip distribution akin to the 
gluonic BM function. In our present Letter, we examine
the diphoton process entirely from the point of view of TMD 
factorization. Focusing on the gluonic Sivers and BM
functions, we restrict ourselves to the application of an effective 
tree-level TMD formalism in the spirit of 
Refs.~\cite{Mulders:1995dh,Arnold:2008kf}. 
At present, we are not able to present a proof that TMD factorization
indeed holds for this process. 
Given the color-singlet nature of the final state and its similarity to DY kinematics, 
it appears plausible that such a factorization could be established
if $Q\sim p_T\gg q_T$, where $Q$ ($q_T$) is the photon pair mass (transverse 
momentum) and $p_T$ 
the transverse momentum of one photon.
We hope that our study will motivate work in this direction. 

Measurements of diphoton production have 
been carried out at the Tevatron~\cite{Acosta:2004sn}. 
Detection of diphoton signals should be well
feasible in polarized $pp$ collisions at RHIC~\cite{bland} -- 
statistics for the reaction will depend of course on the collected 
luminosity. 
Concerning the extraction of TMDs from $pp\to \gamma \gamma X$, a
potential complication arises due to the fact that photons can also 
be produced in jet fragmentation, which would very likely spoil
TMD factorization. Such fragmentation contributions may be 
strongly suppressed or even eliminated by using isolation cuts
on the photons. We leave a more detailed discussion of this issue
to a future publication.

{\it Kinematics.}\,---\,We analyze the diphoton process 
$h(P_a) + h(P_b)\rightarrow \gamma(q_a) + \gamma(q_b) + X$
in the center-of-mass (c.m.) frame of the incoming hadrons 
with momenta $P_a^{\mu}=\sqrt{{S}/{2}}~[ 0\,,1\,,\vec{0}_T]$ and 
$P_b^{\mu}=\sqrt{{S}/{2}}~[ 1\,,0\,,\vec{0}_T]$, where 
we used the light-cone notation $a^\mu=[a^-,a^+,\vec{a}_T]$, 
with $a^{\pm}=(a^{0}\pm a^{3})/\sqrt{2}$, $\vec{a}_{T}=(a^{1},a^{2})$, and 
$S=(P_{a}+P_{b})^{2}$.
The expressions for the photon momenta are simplified for $q_{T}\ll Q$
\cite{Arnold:2008kf}  
\begin{eqnarray}
q_{a(b)}^{\mu}=\tfrac{\sqrt{S}}{2}\big[ x_b\tfrac{1\mp\cos\theta}
{\sqrt{2}}\,,x_a\tfrac{1\pm\cos\theta}{\sqrt{2}}\,,\pm\sqrt{x_ax_b}
\sin\theta \,\vec{e}_{\phi}\big]\, ,
\label{eq:frame}
\end{eqnarray}
where the upper (lower) sign in the above expression 
refers to photon $a$ ($b$), 
$x_{a/b}=q^{2}/(2P_{a/b}\cdot q)$ with 
the photon pair momentum $q=q_{a}+q_{b}$, and 
the spatial orientations of the photons are fixed by their
angles $\theta,\,\phi$ in the Collins-Soper (CS) 
frame~\cite{Collins:1981uk,Arnold:2008kf}. The Lorentz transformation 
between the c.m. frame and CS-frame has been worked out in 
Ref.~\cite{Arnold:2008kf} for the (kinematically identical)
DY process  $hh\rightarrow \ell^+\ell^-X$. 
In Eq.~(\ref{eq:frame}), $\vec{e}_{\phi}=(\cos\phi,\,\sin\phi)$. 
Additional azimuthal 
dependence may be introduced by transverse spin vectors of the hadrons, 
$\vec{S}_{a/bT}=(\cos\phi_{a/b},\,\sin\phi_{a/b})$.
The partonic Mandelstam variables expressed in the c.m.-frame read 
$s=2k_{a}\cdot k_{b}=Q^{2}$, $t=-2k_{a}\cdot q_{a}=-Q^{2}\sin^{2}
\tfrac{\theta}{2}$ and $u=-2k_{b}\cdot q_{a}=-Q^{2}\cos^{2}\tfrac{\theta}{2}$,
where $k_{a/b}$ are the incoming parton momenta.

\begin{figure}[t]
\vspace*{1mm}
\includegraphics[width=8.5cm]{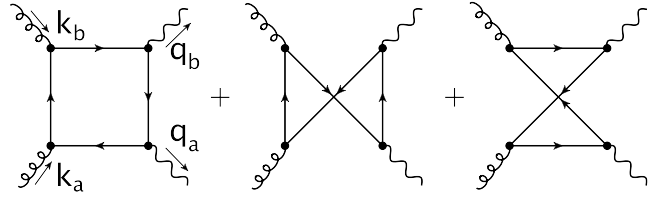}
\caption{Photon pair production by gluon-gluon fusion.\label{fig:QQGG}}
\vspace*{-4mm}
\end{figure}

{\it Photon pair production in $q\bar{q}$ annihilation.}\,---\,At lowest order,
photon pairs are produced through quark-antiquark annihilation, $q\bar{q}\to
\gamma \gamma$. 
By following the same steps of the DY calculation in \cite{Arnold:2008kf}, 
we find for $q_T\ll Q$ 
\begin{equation}
\frac{\d \sigma^{q\qb\rightarrow\gamma\gamma}}{\d^4q\,\d \Omega}\Big|_{q_T\ll Q} = 
\frac{2}{\sin^2\theta}\frac{\d \sigma^{q\qb\rightarrow l^+l^- }}
{\d^4q\,\d \Omega}\Big|_{q_T\ll Q}(e_q^2 \rightarrow e_q^4)\,,
\label{eq:DPQQ}
\end{equation}
where the expression for $\d \sigma^{q\qb\rightarrow l^+l^- }$ can be 
found in Ref.~\cite{Arnold:2008kf}. In Eq.~(\ref{eq:DPQQ}), 
the overall factor $2/\sin^2\theta$ 
is caused by the fact that the process $q\bar{q}\to \gamma \gamma$ proceeds 
via $t$ and $u$ channels while the DY process is via the $s$-channel.
As pointed out above, the diphoton production and DY processes 
share the same quark and antiquark TMDs because both have only initial state interactions. 

{\it Gluon TMDs and photon pair production.}\,---\,Gluon TMDs are defined 
through the correlator~\cite{rodrig}
\begin{eqnarray}
\Gamma_{\mu\nu;\lambda\eta}(x,\vec{k}_{T}) 
& = & \frac{1}{xP^{+}} \int\tfrac{dz^{-}d^{2}z_{T}}{(2\pi)^{3}}\,
\mathrm{e}^{ik\cdot z}\,\label{eq:TMDgluoncorrelator}\\
 &  & \hspace{-1.4cm} \times 
 \langle P,S|\, F_{\mu\nu}^{a}(0)\,\mathcal{W}^{ab}[0\,;\, z]\, 
F_{\lambda\eta}^{b}(z)\,|P,S\rangle\Big|_{z^{+}=0},\nonumber 
\end{eqnarray}
which is a diagonal hadronic matrix element of two field strength 
tensors $F^{\mu\nu}$ between nucleon states with large momentum component 
$P^{+}$ and spin vector $S$. The Wilson line $\mathcal{W}^{ab}$ 
with color indices $a,b$ in the adjoint representation 
makes the correlator gauge invariant~\cite{rogers,tmdfact}.
However, the explicit form of the Wilson line depends on 
the color structure of partonic scattering 
that the gluon TMDs are convoluted with.
For the $gg\to \gamma\gamma$ subprocess of photon pair production, 
we have all gluon TMDs with past-pointing Wilson lines.

The leading terms in a $1/P^{+}$ expansion
of $\Gamma_{\mu\nu;\lambda\eta}$ are given by
$\Gamma^{+i;+j}$ with transverse indices $i,j \in \left\{1,2\right\}$. 
For an unpolarized ($U$)
or a transversely polarized ($T$) hadron of mass $M$ one
has the following decompositions of the correlator $\Gamma^{+i;+j}$ into
gluon TMDs~\cite{rodrig,Meissner:2007rx}:
\begin{eqnarray}
\Gamma_{U}^{+i;+j}(x,\vec{k}_{T}) & = & \tfrac{\delta^{ij}}{2}\, 
f_{1}^{g}+\tfrac{k_{T}^{i}k_{T}^{j}-\frac{1}{2}\vec{k}_{T}^{2}
\delta^{ij}}{2M^{2}}\, h_{1}^{\perp g}\,,\nonumber\\
\Gamma_{T}^{+i;+j}(x,\vec{k}_{T}) & = & -\tfrac{\delta^{ij}}{2}
\tfrac{\epsilon_T^{rs}k_T^r S_T^s}{M}\, f_{1T}^{\perp g}+
\tfrac{i\epsilon_T^{ij}}{2}\tfrac{\kv_T
\cdot\vec{S}_T}{M}g_{1T}^{\perp g}\nonumber\\
 & &\hspace{-2cm}+ \tfrac{S_T^{\{i}\epsilon_T^{j\}r}k_T^r+k_T^{\{i}
\epsilon_T^{j\}r}S_T^r}{8M}h_{1T}^g+\tfrac{k_T^{\{i}\epsilon_T^{j\}r}
k_T^r}{4M^2}\tfrac{\kv \cdot \vec{S}_T}{M}h_{1T}^{\perp g}\,,
\label{eq:GluonTMDs}
\end{eqnarray}
where $\epsilon_T^{ij}\equiv \epsilon^{-+ij}$ and 
$a_T^{\{i}\epsilon_T^{j\}r}\equiv a_T^i\epsilon_T^{jr}+a_T^j\epsilon_T^{ir}$.
Each of the TMDs in~(\ref{eq:GluonTMDs}) is a function of
$x$ and $\vec{k}_{T}^{2}$. The unpolarized correlator $\Gamma_{U}$ 
contains two gluon TMDs, the unpolarized gluon distribution $f_1^g$ and 
the gluonic BM function $h_1^{\perp g}$. 
$\Gamma_T$ is parameterized by the gluon Sivers function $f_{1T}^{\perp g}$, 
and several other gluon TMDs~\cite{rodrig}. Among these, 
$h_{1T}^{\perp g}$ is the gluon ``pretzelosity'' TMD. 

The leading order diagrams for $gg\to\gamma \gamma$ 
are shown in Fig.~\ref{fig:QQGG}. 
Helicity amplitudes for them have been presented in 
Refs.~\cite{Dicus:1987fk,Bern:2001df}. While these could be 
combined in a way suitable for projecting onto transverse
external gluon indices, we choose to compute the diagrams directly,
defining ``semicontracted'' amplitudes $\mathcal{M}_{\pm \pm}^{ij}
\equiv\mathcal{M}_{\,\,\,\,\,\rho\sigma}^{ij}\left(\varepsilon_{\pm}^{\rho}(q_a)
\right)^{*}\left(\varepsilon_{\pm}^{\sigma}(q_b)\right)^{*}$, 
with transverse gluon indices $i,j$, and
contracted with photon polarization vectors $\varepsilon_{\pm}(q_{a/b})$.
To perform the loop integrals, we use standard Mellin-Barnes space methods.
The semicontracted amplitudes $\mathcal{M}_{\pm\pm}^{ij}$ are finite. We obtain 
\begin{eqnarray}
\mathcal{M}_{\pm\pm}^{ik} & = & -\mathcal{K}\left(\varepsilon_{-}^{i}\varepsilon_{-}^{k}+
\varepsilon_{+}^{i}\varepsilon_{+}^{k}-\varepsilon_{\mp}^{i}\varepsilon_{\pm}^{k}+f_{s}\,
\varepsilon_{\pm}^{i}\varepsilon_{\mp}^{k}\right)\,,\nonumber \\
\mathcal{M}_{\pm\mp}^{ik} & = & \mathcal{K}\left(f_{t}\,\varepsilon_{\mp}^{i}\varepsilon_{\mp}^{k}+f_{u}\,
\varepsilon_{\pm}^{i}\varepsilon_{\pm}^{k}+\varepsilon_{-}^{i}\varepsilon_{+}^{k}+\varepsilon_{+}^{i}
\varepsilon_{-}^{k}\right)\,,\label{eq:ResultsHelAmp}
\end{eqnarray}
with $\varepsilon_{\pm}=(1,\pm i)\mathrm{e}^{\mp i\phi}$, $\mathcal{K}\equiv 2\delta^{ab}\alpha_{s}\alpha_{\mathrm{em}} \sum_qe_{q}^{2}$, 
where $e_q$ denotes the fraction of the charge of the quark in the box in units of the elementary charge $e$ and $a$, $b$ are the gluon adjoint color indices. Furthermore, $f_s\equiv L(t,u)$, $f_t\equiv L(s,u)$, 
$f_u\equiv L(s,t)$, with $L$ defined as
\begin{eqnarray}
L(x,y) & = & 1 - \tfrac{x-y}{x+y}\left(\ln |\tfrac{x}{y}|-\i \pi \theta(-\tfrac{x}{y}) \right)\nonumber\\
 & & \hspace{-0.5cm}+\tfrac{1}{2}\tfrac{x^2+y^2}{(x+y)^2}\left(\pi^2 + \left(\ln |\tfrac{x}{y}|-\i \pi 
\theta(-\tfrac{x}{y}) \right)^2 \right)\,.\label{eq:Logs}
\end{eqnarray}
Note that we have $f_s=-M^{(1)}_{++--}$, $f_t=-M^{(1)}_{+-+-}$, $f_u=-M^{(1)}_{+--+}$ 
in terms of the helicity amplitudes of~\cite{Bern:2001df}. 

Our semicontracted amplitudes and the correlator $\Gamma^{+i;+j}$ of 
Eq.~(\ref{eq:TMDgluoncorrelator}) can now be used to compute the cross section for 
$gg\to\gamma \gamma$ in the TMD formalism:
\begin{eqnarray}
\frac{\mathrm{d}\sigma^{gg}}{\mathrm{d}^{4}q\,\mathrm{d}\Omega} & = & \mathcal{H}\int 
d^{2}k_{aT}d^{2}k_{bT}\,\delta^{(2)}(\vec{k}_{aT}+\vec{k}_{bT}-\vec{q}_{T})\times\label{eq:CSTMD1}\\
 &  & \hspace{-1.6cm}\Gamma^{+i;+j}(x_{a},\vec{k}_{aT})\,\Gamma^{-k;-l}(x_{b},\vec{k}_{bT})
\sum_{\lambda_1, \lambda_2}\mathcal{M}^{ik}_{\lambda_1 \lambda_2}
\left(\mathcal{M}_{\lambda_1 \lambda_2}^{jl}\right)^{*}\nonumber\,,
\end{eqnarray}
where $\mathcal{H}=(128(2\pi)^{2}x_{a}x_{b}S^{2})^{-1}$, and where we sum over the photon 
helicities.

Using the decomposition in Eq.~(\ref{eq:GluonTMDs}) we derive from Eq.~(\ref{eq:CSTMD1}) the following result for 
the unpolarized and single transverse spin polarized cross sections in terms of the CS-frame angles:
\begin{eqnarray}
\frac{\mathrm{d}\sigma_{UU}^{gg}}{\mathrm{d}^{4}q\,\mathrm{d}\Omega} & = & \sigma_{0}^{gg}
\Big[\mathcal{F}_{1}(\theta)\,\mathcal{C}\left[f_{1}^{g}\, f_{1}^{g}\right]+\mathcal{F}_{2}(\theta)\,
\mathcal{C}\left[w_{5}\, h_{1}^{\perp g}\, h_{1}^{\perp g}\right]\nonumber \\
 &  & \hspace{-1.5cm}+\cos(2\phi)\Big\{\mathcal{F}_{3}(\theta)\,\left(\mathcal{C}\left[w_{1}\, 
h_{1}^{\perp g}\, f_{1}^{g}\right]+\mathcal{C}\left[w_{2}\, f_{1}^{g}\, h_{1}^{\perp g}\right]\right)\Big\}\nonumber \\
 &  & +\cos(4\phi)\left\{\mathcal{F}_{4}(\theta)\mathcal{C}\left[w_{4}\, h_{1}^{\perp g}\, h_{1}^{\perp g}
\right]\right\} \Big]\,,\label{eq:BMeffect}\\
\frac{\mathrm{d}\sigma_{TU}^{gg}}{\mathrm{d}^{4}q\,\mathrm{d}\Omega} & = & \sigma_0^{gg}\,|\vec{S}_T|
\sin \phi_a\,\Big[\mathcal{F}_1(\theta)\,\mathcal{C}\left[w_3\,f_{1T}^{\perp g}\,f_1^g\right]+\nonumber\\
 & & \hspace{-1.7cm}\mathcal{F}_2(\theta)\,\left(\mathcal{C}\left[w_6\,h_{1T}^g\,h_1^{\perp g}\right]+
\mathcal{C}\left[w_7\,h_{1T}^{\perp g}\,h_1^{\perp g}\right]\right) \,+\, ...\Big]\,,\label{eq:Siveffect}
\end{eqnarray}
where $\sigma_0^{gg}\equiv 2 \mathcal{K}^2 \mathcal{H}$, and where the ellipses denote 
additional terms that vanish upon $\phi$-integration. We have defined 
$\mathcal{F}_{1}(\theta)=f_{s}^{2}+|f_{t}|^{2}+|f_{u}|^{2}+5$, $\mathcal{F}_{2}(\theta)=2(f_{s}-1)$,
$\mathcal{F}_{3}(\theta)=f_{s}+\Re[f_{u}+f_{t}]-1$,
$\mathcal{F}_{4}(\theta)=f_{u}f_{t}^{*}+f_{t}f_{u}^{*}+2$, and
\begin{eqnarray}
\mathcal{C}\left[w\, f_{1}\, f_{2}\right] & \equiv & \int d^{2}k_{aT}d^{2}k_{bT}\,
\delta^{(2)}(\vec{k}_{aT}+\vec{k}_{bT}-\vec{q}_{T})
\nonumber \\
 &  \times &
w(\vec{k}_{aT},\vec{k}_{bT}) f_{1}(x_{a},\vec{k}_{aT}^{2})
f_{2}(x_{b},\vec{k}_{bT}^{2}).
\label{eq:TMDnotation}
\end{eqnarray}
Defining $(ab)_\pm
\equiv (a_1 b_1 \pm a_2 b_2)/2M^2$ and 
$[ab]_{\pm}\equiv (a_1 b_2 \pm a_2 b_1)/2M^2$,
the following weights appear in \eqref{eq:BMeffect} and \eqref{eq:Siveffect}:
\begin{eqnarray}
 &  w_1=-2 (k_{aT} k_{aT})_-\,\, ,\,\, 
w_2=-2 (k_{bT} k_{bT})_- \,\,, & \nonumber\\
& w_3=\frac{1}{M}k_{aT,2}\,\, , \, \  w_4 = (k_{aT}k_{bT})_-^2 - [k_{aT} k_{bT}]_+^2 \,\,,\,\, & \nonumber \\
& w_5 =   [k_{aT} k_{bT}]_-^2 -(k_{aT}k_{bT})_+^2 \,\,,\,\, & \nonumber\\
& w_6 =  \frac{1}{2M}\left( (k_{bT}k_{bT})_+ k_{aT,2}-2 (k_{aT} k_{bT})_+  
k_{bT,2} \right) \,\,,\,\, & \nonumber\\
 & w_7=-\frac{2}{M} (k_{aT}k_{bT})_+ [k_{aT}k_{bT}]_- k_{aT,1} \; .&
\label{eq:ws}
\end{eqnarray}
We stress that the angular structure of the unpolarized cross section shown in~(\ref{eq:BMeffect}) 
is identical to that found in the context of collinear factorization for perturbative soft-gluon 
radiation from the LO process $gg\to\gamma\gamma$~\cite{Nadolsky,Catani}. This may hint at a possible 
matching of the TMD and collinear formalisms in the intermediate $q_T$ region $\Lambda_{\mathrm{QCD}}
\ll q_T\ll Q$. We also note that weighted cross sections of the form
$\langle F\rangle\equiv\int d^{2}q_{T}\, d\phi\, F(q_{T},\phi)\,
(\mathrm{d}\sigma/\mathrm{d}^{4}q\,\mathrm{d}\Omega)$
may help in disentangling the various terms in 
~(\ref{eq:BMeffect}) and (\ref{eq:Siveffect}). For instance,
$\langle q_{T}^{4}\cos(4\phi)\rangle\propto h_{1}^{\perp(2)g}(x_{a})\, h_{1}^{\perp(2)g}(x_{b})$, and 
$\langle q_{T}^{2}\cos(2\phi)\rangle\propto h_{1}^{\perp(2)g}(x_{a})\, f_1^{(0)g}(x_{b})$, with $k_T$ moments of $f_1^g$ and $h_1^{\perp g}$.

{\it IV. Numerical estimates.}\,---\,In 
order to estimate the size of the various contributions to~\eqref{eq:BMeffect},\eqref{eq:Siveffect}, 
we use a Gaussian model for the TMDs as frequently chosen for the analysis of SIDIS 
or DY data~\cite{Barone:2010zz,Schweitzer:2010tt}. For the unpolarized quark and gluon TMDs 
$f_{1}^{q,g}$ we make the ansatz
\begin{eqnarray}
\hspace{-0.4cm}f_{1}^{q}(x,\vec{k}_{T}^{2})=\tfrac{f_{1}^{q}(x)}{\pi\beta}\,\mathrm{e}^{-\tfrac{\vec{k}_{T}^{2}}{\beta}} 
& , & f_{1}^{g}(x,\vec{k}_{T}^{2})=\tfrac{G(x)}{\pi\gamma}\,\mathrm{e}^{-\tfrac{\vec{k}_{T}^{2}}{\gamma}},
\label{eq:GaussianAnsatz}
\end{eqnarray}
with widths $\beta$ and $\gamma$ for which we assume 
$\beta=\gamma=0.5\,\mathrm{GeV^{2}}$ at RHIC, 
and with the $k_T$-integrated parton distributions of~\cite{cteq6m}.
\begin{figure}[t]

\includegraphics[width=5.5cm,angle=270]{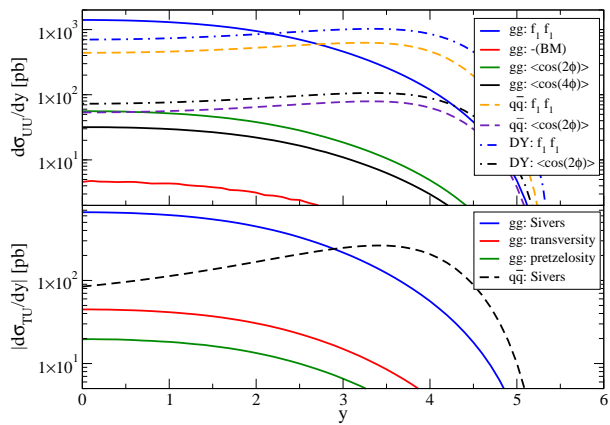}
\caption{Pair rapidity ($y$) dependence of the various terms in the cross 
sections in Eqs.~\eqref{eq:BMeffect} and \eqref{eq:Siveffect}, 
for the unpolarized [top] 
and single transversely polarized [bottom] cases, in $pp$ collisions at
$\sqrt{S}=500$~GeV. For the spin-dependent cross section we show the absolute value 
since the sign of the TMDs is not fixed by the positivity bounds.
For comparison we also show 
predictions for the unpolarized Drell-Yan process in the upper panel,
without any cuts on lepton transverse momenta. }
\label{fig:The--dependence-at}
\end{figure}
Very little is known about the other quark and gluon TMDs 
at RHIC energies.
Model-independent positivity bounds 
for them were derived in Refs.~\cite{rodrig,Bacchetta:1999kz}. To estimate 
the maximally possible effects in the diphoton process we assume saturation of 
these positivity bounds for both quarks and gluons.
For the gluon Sivers function this gives approximately 
\begin{eqnarray}
|f_{1T}^{\perp g}| \simeq \tfrac{M}{k_T} f_1^{g} \,.\label{eq:PosBgl}
\end{eqnarray}
Similarly the positivity bounds lead to the following approximations for the other TMDs:
$|h_1^{\perp g}|\simeq (2M^2)/k_T^2\,f_1^g$,  $|h_1^{\perp q}|\simeq |f_{1T}^{\perp q}| \simeq M/k_T\,f_1^q$,  
$|h_1^{g}|\simeq M/k_T\,f_1^g$ (with $h_1^{g}=h_{1T}^g+k_T^2/(2M^2)
h_{1T}^{\perp g}$), and $|h_{1T}^{\perp g}|\simeq (2M^3)/k_T^3\,f_1^g$.

In Fig.~\ref{fig:The--dependence-at} we present our numerical estimates 
from our Gaussian ansatz.  In generating those curves, we required each photon to have
a transverse momentum of at least 1~GeV, and we integrated over $4\leq Q^2\leq 30$~GeV$^2$, 
$0\leq q_T\leq 1$~GeV, and the CS angles with appropriate azimuthal 
weightings.  For the unpolarized cross section (upper panel), 
the $gg\to\gamma\gamma$ channel dominates at midrapidity 
while the $q\bar{q}\to\gamma\gamma$  channel is more important 
at forward/backward rapidity ($|y|>2$) of the photon pair. 
The contribution by the gluon BM effect to 
the $\phi$-{\it in}dependent cross section turns out to be rather small. 
On the other hand, the $\cos2\phi$ and $\cos4\phi$ contributions 
induced by gluons could be at the percent level 
for TMDs saturated by the positivity bounds.  Realistically, however, 
one may expect smaller effects depending on the actual size of the TMDs.
In order to estimate the maximum size of quark Sivers contribution to the spin-dependent cross section, we kept only the positivity bound saturated up-quark Sivers function since the up- and down-quark Sivers functions have an opposite sign. 
From the lower panel in Fig.~\ref{fig:The--dependence-at}, it is important to note that  
the gluon Sivers effect exceeds the quark Sivers effect 
by a factor seven or so 
at midrapidity, and dominates the contribution for a wide range of rapidity.  That is, the single transverse spin asymmetry of the diphoton production at RHIC could offer excellent opportunities for exploring the gluon Sivers function.  
Other effects caused by the gluon TMDs $h_{1T}^g$ and $h_{1T}^{\perp,g}$ are negligible.

{\it V. Conclusion.}\,---\,
We have investigated photon pair production in hadronic collisions in the framework of 
TMD factorization. We have shown that this process may be suited for studying gluon TMDs at RHIC. 
The $\cos(4\phi)$ modulation can be used to extract the gluon Boer-Mulders function. Even a small effect can be significant since this modulation is absent in the $q\bar{q}$ channel. The $\cos(2\phi)$ modulation ultimately gives information on the sign of $h_{1}^{\perp g}$. Such measurements may also be performed at the LHC where 
the production rate from gluon fusion is much larger.
Another unique feature of the diphoton process is its sensitivity to 
the gluon Sivers function in polarized proton collisions. 
Measurements at RHIC could hence provide 
important clues about the correlation between gluon motion and hadron spin.

We thank L.~Bland, D.~Boer, D.~de Florian and A.~Metz for helpful discussions.
This work has been supported by the U.S. Department of Energy (Contract 
No. DE-AC02-98CH10886).



\begin{thebibliography}{99}
%

\bibitem{sivers} D.~W.~Sivers,
  Phys.\ Rev.\  D {\bf 41}, 83 (1990);
  Phys.\ Rev.\  D {\bf 43}, 261 (1991).

\bibitem{bm} 
  D.~Boer and P.~J.~Mulders,
  Phys.\ Rev.\  D {\bf 57}, 5780 (1998).

\bibitem{Barone:2010zz} For review, see:
  V.~Barone, F.~Bradamante and A.~Martin,
  Prog.\ Part.\ Nucl.\ Phys.\  {\bf 65}, 267 (2010).

\bibitem{Collins:1981uk} J.~C.~Collins and D.~E.~Soper,
  Nucl.\ Phys.\  B {\bf 193}, 381 (1981)
  [Erratum-ibid.\  B {\bf 213}, 545 (1983)].

\bibitem{tmdfact}
X.~D.~Ji, J.~P.~Ma and F.~Yuan,
  Phys.\ Rev.\  D {\bf 71}, 034005 (2005);
  Phys.\ Lett.\  B {\bf 597}, 299 (2004);
 J.~C.~Collins and A.~Metz,
  Phys.\ Rev.\ Lett.\  {\bf 93}, 252001 (2004);
 J.~C.~Collins, T.~C.~Rogers and A.~M.~Stasto,
  Phys.\ Rev.\  D {\bf 77}, 085009 (2008).

\bibitem{rodrig}
P.~J.~Mulders and J.~Rodrigues,
  Phys.\ Rev.\  D {\bf 63}, 094021 (2001);
  X.~d.~Ji, J.~P.~Ma and F.~Yuan,
  JHEP {\bf 0507}, 020 (2005).

\bibitem{boerwv} D.~Boer and W.~Vogelsang,
  Phys.\ Rev.\  D {\bf 69}, 094025 (2004);
A.~Bacchetta {\it et al.},
  Phys.\ Rev.\ Lett.\  {\bf 99}, 212002 (2007).

\bibitem{gluoncharm} F.~Yuan,
  Phys.\ Rev.\  D {\bf 78}, 014024 (2008);
Z.~B.~Kang and J.~W.~Qiu,
  Phys.\ Rev.\  D {\bf 78}, 034005 (2008);
F.~Yuan and J.~Zhou,
  Phys.\ Lett.\  B {\bf 668}, 216 (2008);
Z.~B.~Kang {\it et al.},
  Phys.\ Rev.\  D {\bf 78}, 114013 (2008);
  M.~Anselmino {\it et al.},
  Phys.\ Rev.\  D {\bf 70}, 074025 (2004).

\bibitem{Boer:2010zf}
  D.~Boer {\it et al.},
Phys.\ Rev.\ Lett.\  {\bf 106}, 132001 (2011).

\bibitem{Boer:2009nc}
D.~Boer, P.~J.~Mulders and C.~Pisano,
  Phys.\ Rev.\  D {\bf 80}, 094017 (2009).

\bibitem{rogers} J.~C.~Collins and J.~W.~Qiu,
  Phys.\ Rev.\  D {\bf 75}, 114014 (2007);
 T.~C.~Rogers and P.~J.~Mulders,
  Phys.\ Rev.\  D {\bf 81}, 094006 (2010);
C.~J.~Bomhof, P.~J.~Mulders and F.~Pijlman,
  Phys.\ Lett.\  B {\bf 596}, 277 (2004).

\bibitem{Berger}
  E.~L.~Berger, E.~Braaten and R.~D.~Field,
  Nucl.\ Phys.\  B {\bf 239}, 52 (1984).

\bibitem{Nadolsky} C.~Balazs {\it et al.},
  Phys.\ Lett.\  B {\bf 637}, 235 (2006); P.M.~Nadolsky {\it et al.},
  Phys.\ Rev.\  D {\bf 76}, 013008 (2007);
  C.~Balazs {\it et al.}, Phys.\ Rev.\ D{\bf 76}, 013009 (2007).

\bibitem{Catani} S.~Catani and M.~Grazzini,
  Nucl.\ Phys.\  B {\bf 845}, 297 (2011).

\bibitem{Mulders:1995dh} P.~J.~Mulders and R.~D.~Tangerman,
  Nucl.\ Phys.\  B {\bf 461}, 197 (1996)
  [Erratum-ibid.\  B {\bf 484}, 538 (1997)];
A.~Bacchetta {\it et al.},
  JHEP {\bf 0702}, 093 (2007);

\bibitem{Arnold:2008kf}
  S.~Arnold, A.~Metz and M.~Schlegel,
  Phys.\ Rev.\  D {\bf 79}, 034005 (2009).


\bibitem{Acosta:2004sn}
  D.~E.~Acosta {\it et al.}  [CDF Collaboration],
  Phys.\ Rev.\ Lett.\  {\bf 95}, 022003 (2005);
  V.~M.~Abazov {\it et al.}  [The D0 Collaboration],
  Phys.\ Lett.\  B {\bf 690}, 108 (2010).

\bibitem{bland} Leslie Bland, {\it private communication}.

\bibitem{Meissner:2007rx} For notation, see:
S.~Meissner, A.~Metz and K.~Goeke,
  Phys.\ Rev.\  D {\bf 76}, 034002 (2007).

\bibitem{Dicus:1987fk}
  D.~A.~Dicus and S.~S.~D.~Willenbrock,
  Phys.\ Rev.\  D {\bf 37}, 1801 (1988).

\bibitem{Bern:2001df}
  Z.~Bern, A.~De Freitas and L.~J.~Dixon,
  JHEP {\bf 0109}, 037 (2001).


\bibitem{Schweitzer:2010tt}
  For review, see: P.~Schweitzer, T.~Teckentrup and A.~Metz,
  Phys.\ Rev.\  D {\bf 81}, 094019 (2010).

\bibitem{cteq6m} J.~Pumplin {\it et al.},
  JHEP {\bf 07}, 012 (2002).

\bibitem{Bacchetta:1999kz}
  A.~Bacchetta {\it et al.},
  Phys.\ Rev.\ Lett.\  {\bf 85}, 712 (2000).

\end{thebibliography}
\end{document}